\documentstyle[eqsecnum,aps,prd]{revtex}
\input epsf
\begin{document}

\preprint{MADPH-99-1151}

\wideabs{ 
\title{From scalar to string confinement} 
\author{Theodore J. Allen}
\address{School of Arts \& Sciences, SUNY Institute of Technology \\
P.O. Box 3050, Utica, New York 13504 USA}

\author{M. G. Olsson}
\address{Department of Physics, University of Wisconsin, \\
1150 University Avenue, Madison, Wisconsin 53706 USA }

\author{Sini\v{s}a Veseli}
\address{Fermi National Accelerator Laboratory \\
P.O. Box 500, Batavia, Illinois 60510 USA}

\date{\today}
\maketitle

\begin{abstract}
We outline a connection between scalar quark confinement, a
phenomenologically successful concept heretofore lacking fundamental
justification, and QCD.  Although scalar confinement does not follow from
QCD, there is an interesting and close relationship between them. We
develop a simple model intermediate between scalar confinement and the QCD
string for illustrative purposes.  Finally, we find the bound state masses
of scalar, time-component vector, and string confinement analytically
through semi-classical quantization.
\end{abstract}
}  

\section{Introduction}\label{sec:intro}

Going beyond the non-relativistic potential model of quark confinement
means that more than the static interaction energy must be specified.  In
the language of potential models the Lorentz nature of the interaction is
needed.  To agree with the observed spin-orbit splitting it was proposed
long ago \cite{ref:one} that the large distance (confining) potential is a
Lorentz scalar.  In this case there is no magnetic field to influence the
quarks' spins and the only spin-orbit interaction is the kinematic ``Thomas
term.''  The Thomas type spin-orbit interaction partially cancels that of
the short range one-gluon exchange, in agreement with the observed
spectrum.

Some insight into the use of the scalar potential was given by Buchm\"uller
\cite{ref:two}.  His argument is that at large distances one expects the
QCD field of the quarks to become string- or flux-tube-like.  The QCD flux
tube is purely chromoelectric in its rest frame, and hence in the rest
frame of each quark there is no chromomagnetic field to provide a
spin-orbit interaction.  The scalar interaction yields this same result by
fiat; there is no magnetic field anywhere because it is not a vector-type
interaction.  This provides some justification for using the scalar
potential but does not establish a direct connection. It agrees only in
having the same spin-orbit interaction at long range as QCD.

Subsequently it was shown that for slowly moving quarks, QCD predicts both
spin-dependent\cite{ref:three} and spin-independent\cite{ref:four}
relativistic corrections. The long-range spin dependence is just the Thomas
type spin-orbit interaction.  The spin-independent corrections differ from
those of scalar confinement \cite{ref:five,ref:six}.  It also has been
established that the QCD predictions at long distance are the same as those
of a string or flux tube interaction \cite{ref:six}.  Lattice simulations
also favor the Thomas interaction \cite{ref:seven}.

Since spin-independent effects are difficult to identify from the data,
scalar confinement remains phenomenologically successful.  As scalar
confinement is also relatively simple computationally, it continues to be a
popular and useful tool in hadron physics.  It should be pointed out that
its use in the Salpeter equation leads to cancellations \cite{ref:eight} in
the ultra-relativistic limit, resulting in a very non-linear Regge
trajectory \cite{ref:eight,ref:nine,ref:ten}.  

Although scalar confinement has been used for a long time in hadron
physics, its relation to QCD has never been clarified.  It is the purpose
of this paper to place scalar confinement in relation to QCD and in
particular to the QCD string.  In section \ref{sec:two} we point out that
there is a certain four-vector potential that is isomorphic to a scalar
potential.  In section \ref{sec:three} we compare this four-vector
potential to the QCD string.  Noting certain similarities and differences,
we propose a model intermediate between the string and scalar confinement.
The semi-relativistic reductions for scalar, time-component vector,
intermediate, and string confinements are compared in section
\ref{sec:four}.  Although by construction, all these confinement models
have the same non-relativistic limit, their relativistic reductions differ.
In section \ref{sec:five} we explore the ``ultra-relativistic'' Regge
sector with a massless quark via semi-classical quantization.  The Regge
behavior of the different confinement models show some remarkable
similarities and differences.  Finally, in section \ref{sec:six} we present
our conclusions and summarize our work.

\section{The four-vector potential isomorphic to the scalar potential}
\label{sec:two}

The action for a scalar (spinless) quark moving in Lorentz scalar and
four-vector potentials, $\phi(x)$, and $A_\mu(x)$ respectively, is
\begin{equation}
S = -\int d\tau\, \left[m + \phi(x) - u^\mu A_\mu(x)\right] \ ,
\label{eq:genact}
\end{equation}
where $m$ is the rest-mass of the quark, $u^\mu$ is the quark's
four-velocity, and $d\tau$ is the proper time element $dt/\gamma$.  The
quark four-velocity, $u^\mu = (\gamma,\gamma{\bf v})$, with $ \gamma =
(1-{\bf v}^2)^{-1/2}$, satisfies $-u^\mu u_\mu = 1$.

When $A_\mu(x) \equiv 0$, the action (\ref{eq:genact}) reduces to the usual
scalar potential action.  On the other hand, when $\phi(x) = 0$, the action
(\ref{eq:genact}) describes a quark moving in an ``electromagnetic''
($U(1)\subset SU(3)_{\rm color}$) color field.  It was pointed out by
Buchm\"uller \cite{ref:two} that in the rest frame of the QCD flux tube
there is no color magnetism so that the only spin orbit interaction is
Thomas precession.  If we want to implement Buchm\"uller's criterion we may
assume \cite{ref:eleven} that in the quark rest frame
\begin{equation}\label{eq:Aprime}
A^{\mu^\prime}(x) = (\phi(r), {\bf 0})\ ,
\end{equation}
where $\phi(r) \equiv A^{0^\prime}(x)$ is the time component of
$A^{\mu^\prime}(x)$.  In the laboratory frame, where the quark velocity is
${\bf v}$, the four-vector potential is
\begin{equation}
A^\mu = u^\mu\phi(r) = (\gamma,\gamma{\bf v})\phi(r)\ .
\end{equation}
We note that the components depend on both position and velocity.  The
vector potential contributes to the action (\ref{eq:genact}) as
\begin{equation}
u^\mu A_\mu = - \phi(r)\ .
\end{equation}
The resulting contribution is exactly the same as the scalar potential in
Eq.~(\ref{eq:genact}).  The four-vector potential corresponding to $\phi(r) =
ar$ was discussed by us earlier in Ref.~\cite{ref:eleven}.

By this simple demonstration we have shown that there are two Lorentz type
potentials that have identical consequences.  The four-vector version is
apparently more closely related to QCD.  As we will see, we can quite
closely draw similarities and differences.

\section{Comparing scalar and string confinement - an intermediate model
emerges}\label{sec:three} 

For a spinless quark moving relative to a heavy quark at the origin, the
action can be written as the time integral of a function of the light
quark's position and velocity,
\begin{equation}
S = \int dt\, L({\bf r}, {\bf v})\ .
\end{equation}

If we consider the quark as a particle of mass $m$ moving in a linear
scalar confining potential $\phi(r)= ar$, its Lagrangian is
\begin{eqnarray}\label{eq:scalarlag}
L_{\rm scalar} & = & -\,\gamma^{-1}\left(m+\phi(r)\right) \nonumber \\
               & = & -m\sqrt{1 - v^2} - ar \sqrt{1 - v^2}\ .
\end{eqnarray}

At large distances, QCD is thought to resemble a Nambu-Goto string or flux
tube model.  For a scalar quark at the end of a straight flux tube, the
corresponding Lagrangian is \cite{ref:twelve}
\begin{equation}\label{eq:fluxstring}
L_{\rm string} = -m\sqrt{1-v^2} - ar \int_0^1 d\sigma\, \sqrt{1 -
\sigma^2 v_\perp^2}\ ,
\end{equation}
where $v_\perp$ is the quark velocity transverse to the string.  Comparing
the scalar and string interactions, we see there are two evident
differences.  The first is that the string energy is spread along the
length of the string whereas in the scalar potential case the energy may be
thought of as being concentrated at the quark coordinate.  The second is
that because of the reparametrization invariance of the Nambu-Goto action
(which physically is the invariance of an electric field to boosts along
its direction), from which Eq.~(\ref{eq:fluxstring}) follows, only the
transverse velocity of the string may appear in the interaction energy.

The first distinction can be considered as a quantitative one which leaves
the basic structure unchanged.  This difference changes the velocity
dependence of the additional three-momentum due to the interaction from
${\bf p} = a r {\bf v}$ in the scalar case to ${\bf p} = {ar\over 2v_\perp}
\left[{\arcsin v_\perp\over v_\perp} - \sqrt{1-v_\perp^2}\right]\hat{{\bf
v}}_\perp$ for the string.

The second distinction has far-reaching consequences.  In a non-rotating
($s$-wave) system, the scalar interaction contributes to the momentum
whereas the string does not. The string Hamiltonian contributes only as the
time component of a vector potential (vector-like) while the scalar
Hamiltonian remains scalar. 

It is instructive to construct a confinement model in which one of the
above distinctions is removed.  We will briefly consider the intermediate
model having Lagrangian
\begin{equation}\label{eq:intermed}
L_{\rm Int} = -m\sqrt{1-v^2} - ar\sqrt{1 - v_\perp^2}\ .
\end{equation}
We note that although the interaction is concentrated at the quark
position, it depends only on the transverse velocity.

This Lagrangian will lead to a Hamiltonian having characteristics of
the string while remaining algebraically tractable.

In the usual way, the Hamiltonian corresponding to Eq.~(\ref{eq:intermed})
is found to be
\begin{equation}\label{eq:HamInt}
H_{\rm Int} = m\gamma + ar\gamma_\perp\ ,
\end{equation}
and the angular momentum, $J = \partial L_{\rm Int} / \partial \omega$,
with $v_\perp = \omega r$, is
\begin{equation}\label{eq:AngMomInt}
J = m\gamma v_\perp r + a r^2 \gamma_\perp v_\perp\ .
\end{equation}

Unlike in the string system, the velocities here can be eliminated in favor of
the momenta, making this model much more tractable.  From the definition of
radial momentum
\begin{equation}
p_r = {\partial L_{\rm Int}\over \partial \dot{r}} = m\gamma\, \dot r\ ,
\end{equation}
the useful identity 
\begin{equation}
m\gamma  =  W_r \gamma_\perp\ , \label{eq:useful}
\end{equation}
with
\begin{equation}
W_r   \equiv  \sqrt{p_r^2 + m^2}\ , \label{eq:Wr}
\end{equation}
follows.

Using the identity (\ref{eq:useful}), we find that $H_{\rm Int}$ and $J$ of
Eqs.~(\ref{eq:HamInt}), (\ref{eq:AngMomInt}) become
\begin{eqnarray}
H_{\rm Int} & = & (W_r + ar) \gamma_\perp\ , \label{eq:Hint}\\
J & = & r v_\perp \gamma_\perp (W_r + ar)\ . \label{eq:Jint} 
\end{eqnarray}
We can solve Eq.~(\ref{eq:Jint}) for $\gamma_\perp$ using $v_\perp^2 = 1 -
\gamma_\perp^{-2}$, and substituting into Eq.~(\ref{eq:Hint}) to obtain
\begin{equation}\label{eq:Hint2}
H_{\rm Int} = \sqrt{{J^2\over r^2} + (W_r + ar)^2} \ . 
\end{equation}

\section{Comparing relativistic corrections of spinless confinement
models}\label{sec:four} 

As we have seen, there are several types of confinement models, even for
spinless quarks.  In this section we will enumerate and compare the
relativistic reductions of various models.  We first consider the
relativistic reductions of the classic static potential models.

\subsection{Scalar confinement}	

From the scalar interaction Lagrangian (\ref{eq:scalarlag}) with $\phi =
ar$, we find the canonical three-momentum to be
\begin{equation}
{\bf p} = (m+ar)\gamma  {\bf v}\ ,
\end{equation}
which results in the Hamiltonian
\begin{equation}\label{eq:scalarham}
H = \sqrt{p^2 + (m+ar)^2} \ .
\end{equation}
For $m \gg ar$ and $m \gg p$, we expand to obtain the
relativistic corrections
\begin{eqnarray}
H & \simeq & \sqrt{p^2 + m^2} + ar - {a\over 2m^2} p^2 r +
\ldots\ \nonumber
\\
& = & \sqrt{p^2 + m^2} + ar - {a p_r^2 r\over 2m^2}  - {a J^2 \over 2m^2r}+
\ldots\ .\label{eq:Hscalar}
\end{eqnarray}
Even though scalar confinement will yield, for spin-1/2 quarks, the
spin-orbit interaction consistent with experiment, lattice, and QCD, the
spin-independent terms in Eq.~(\ref{eq:Hscalar}) are inconsistent with QCD
\cite{ref:five,ref:six}.

\subsection{Time component vector confinement}

In time component vector confinement models, the potential $ar$ is taken to
be the (laboratory frame) time component of a vector potential $A^\mu$;
$A^\mu = (ar, {\bf 0})$.  The quark Lagrangian then is
\begin{equation}
L_{\rm vector} = -m\sqrt{1-v^2} - ar\ .
\end{equation}
The canonical three-momentum following from this Lagrangian, 
\begin{equation}
{\bf p} = \nabla_{\bf v} L = m\gamma {\bf v}\ ,
\end{equation}
leads to the Hamiltonian
\begin{equation}\label{eq:Hvector}
H  = \sqrt{m^2 + p^2} + ar\ .
\end{equation}
There are no relativistic corrections other than kinetic energy
corrections. Vector confinement is disfavored since the associated
spin-orbit interaction adds to the short range spin-orbit interaction
giving spin-orbit splittings that are too large when compared to
experimental values or lattice simulations.

\subsection{Intermediate model}

The Hamiltonian for this model was given in Eq.~(\ref{eq:Hint2}).  The
relativistic reduction for $m \gg ar$ and $m \gg p$ is
\begin{equation}
H_{\rm Int} \simeq \sqrt{p^2 + m^2} + ar  - {a J^2 \over 2m^2r}+
\ldots\ .
\end{equation}
Comparing to $H_{\rm scalar}$ in Eq.~(\ref{eq:Hscalar}), we see the same
reduction except for the missing $p_r$ term.  This might be expected since
the interaction does not contribute to the radial momentum.  We discuss
this result further in the following subsection.

\subsection{String confinement}

The reduction of the string is discussed in Ref.~\cite{ref:six}, where it
was shown that the string contributes a rotational energy equal to that of
a uniform rod of length $r$ and mass $ar$.  This energy is
\begin{eqnarray}
E_R & = & \frac12 I \omega^2 \ , \\
& = & \frac12 k (ar) r^2 \left({J\over mr^2}\right)^2 \ , \\
&=& {k a J^2 \over 2 m^2 r}\ ,
\end{eqnarray}
where the geometrical factor $k = \frac13$ for a uniform rod.  If all of
the ``mass'' of the string is concentrated at the position of the moving
quark end, then $k=1$.

The ``kinetic'' energy term, when expanded, yields
\begin{equation}
\sqrt{p^2 + m^2} \simeq m + {p^2\over 2m} - {p^4\over 8 m^3} + \ldots \ .
\end{equation}
In the semi-relativistic regime the momentum is mostly that of the quark
with a small contribution from the ``interaction.''
\begin{eqnarray}
p^2 & = & p_r^2 + {1\over r^2} (J_q + J_{\rm in})^2\ \nonumber \\
    & \simeq & \left(p_r^2 + {J_q^2 \over r^2}\right) + {2J_q J_{\rm in}\over
r^2}\, \nonumber \\ 
& \simeq & p_q^2 + {2J_q\over r^2} \left({J_q\over
mr^2}\right) \left(k a r^3\right) \ , \\
p^2 & \simeq & p_q^2 + 4m  E_R\ ,
\end{eqnarray}
and hence, 
\begin{equation}
\sqrt{p^2 + m^2 } \simeq \sqrt{p_q^2 + m^2 } + 2 E_R \ . 
\end{equation}
So, if one separates the Hamiltonian into the quark's energy plus an
interaction energy $ar + E_R$, then
\begin{equation}
H \simeq\sqrt{p^2 + m^2 }+ ar - E_R \ .
\end{equation}
This is exactly what one finds in the intermediate models with $k=1$.  The
string Hamiltonian is then the same, only with $k=\frac13$.
\begin{equation}
H \simeq \sqrt{p^2 + m^2 }+ ar - {a J^2 \over 6 m^2 r} \ .
\end{equation}
This result follows systematically from the string invariants
(\ref{eq:stringJ}) and (\ref{eq:stringH}) in the large mass expansion
\cite{ref:six}.  

\section{Comparing Regge Structures of Spinless Confinement
Models}\label{sec:five} 

In this section we explore both the analytic and the numerical solutions
for the Regge spectroscopy expected from the previously considered models.
In particular, we investigate the ultra-relativistic limit when the
``light'' quark has zero mass.  The extension to two light quarks is
straightforward.  It is in this ``massless'' limit where straight Regge
trajectories with evenly spaced daughter trajectories are obtained in many
confinement models and a close correspondence to observed light and
heavy-light mesons is expected.  In our analytical work we will usually
assume that the orbital excitations are large compared to the radial
excitation.  We may consequently expect the semi-classical quantization
scheme to be quite accurate.  Quantization is carried out by performing the
phase-space integral,
\begin{equation}
2\pi(n+\Gamma) = \oint p_r\, dr = 2\int_{r_-}^{r_+} p_r\, dr\ , 
\end{equation}
where $r_{\pm}$ are the classical turning points and $\Gamma$ is a constant
that depends upon the problem.\footnote{Roughly, $\Gamma$ depends on the
nature of the potential at the turning point.  For two smooth turning
points $\Gamma=\frac12$, and for two rigid walls $\Gamma=1$.  For the mixed
case of one of each, $\Gamma = \frac34$.}  As shown by Langer
\cite{ref:thirteen}, the classical angular momentum $J$ must be replaced by
$J+\frac12$ in the expression for the radial momentum $p_r$.

In all cases considered here, the quantization integral can be written, or
accurately approximated by,
\begin{equation}\label{eq:GenInt}
\int_{r_-}^{r_+} p_r \ dr  = C\int_{y_-}^{y_+} {dy\over y}\ \sqrt{(y_+ -
y)(y-y_-)} \ ,
\end{equation}
where $y$ is either $r$ or $r^2$ and $C$ is a constant.  This integral can
be carried out to yield the the semi-classical quantization relation
\begin{equation}\label{eq:GenSemiClass}
n+\Gamma = {C\over 2}\left[\strut y_+ + y_- - 2\sqrt{y_+y_-}\,\right]\ .
\end{equation}

\subsection{Scalar confinement}

We first consider the scalar case because of its simplicity and its central
role in this paper.  The square scalar Hamiltonian, (\ref{eq:Hscalar}),
with the light quark massless is
\begin{equation}
H^2 = p^2 + a^2 r^2\ .\label{eq:Hscalar1}
\end{equation}
This is equivalent to the three-dimensional harmonic oscillator and its
eigenvalues are well-known to be 
\begin{equation}\label{eq:ExactHO}
M^2 = 2a\left(J + 2n + \frac32\right)\ ,\quad J,n = 0,1,2,3,\ldots\ ,
\end{equation}
where $J$ is now the angular momentum quantum number.  To connect with the
analytic solutions to the remaining confinement models we compute the
semi-classical solution for this interaction.

Semi-classical quantization starts with the separation of the momentum into
angular and radial pieces, $p^2 = p_r^2 + {J^2\over r^2}$, and hence
\begin{equation}
p_r^2 = M^2 - {J^2\over r^2} - a^2 r^2\ .
\end{equation}
The classical turning points ($p_r = 0$) satisfy
\begin{eqnarray}
r^2_+ + r^2_- & = & \left({M\over a}\right)^2\ , \nonumber \\
r_+ r_-  & = & {J\over a} \ , \\
{a\over r} \sqrt{(r_+^2 - r^2)(r^2-r_-^2)}& = & p_r \ .\nonumber
\end{eqnarray}
Comparing this last relation to Eq.~(\ref{eq:GenInt}), we read off
$C=\frac{a}2$, and $y=r^2$, and by Eq.~(\ref{eq:GenSemiClass}) with $\Gamma =
\frac12$ and $J\rightarrow J+\frac12$, we find
\begin{equation}
n + \frac12 = {a\over 4}\left[{M^2\over a^2} - {2\over
a}\left(J+\frac12\right)\right]\ ,
\end{equation}
which yields
\begin{equation}\label{eq:ScalarE2}
M^2 = 2a\left[J + 2n + \frac32\right]\ ,
\end{equation}
identical to the exact solution (\ref{eq:ExactHO}).

\begin{figure}[hbp]
\epsfxsize = \linewidth 
\hspace*{-5mm}
\epsfbox{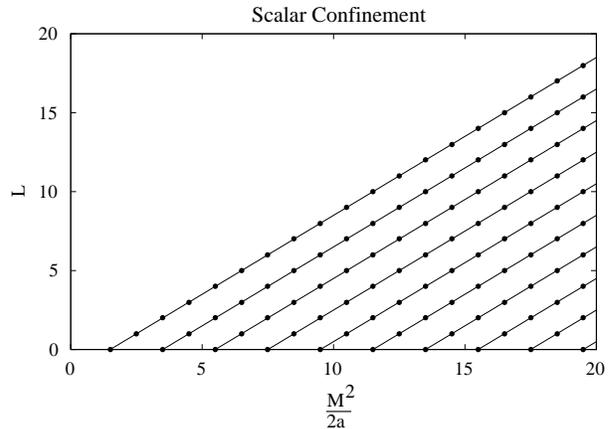}
\vskip 1 cm
\caption{Regge structure and states in pure linear scalar confinement from
numerical diagonalization of the Hamiltonian (\protect\ref{eq:scalarham}) with
$m=0$.  Solid lines are the semi-classical result, which is exact for the
squared Hamiltonian (\protect\ref{eq:Hscalar1}).}
\label{fig:one}
\end{figure}

In Fig.~\ref{fig:one} we show the Regge plot for pure scalar confinement.
The dots represent the exact numerical solution by the variational method,
for instance see the appendix in Ref.~\cite{ref:eight}.  The numerical
solutions correspond to the unsquared Hamiltonian (\ref{eq:scalarham}) with
$m=0$. The lines are the analytic solution, Eq.~(\ref{eq:ExactHO}) or
(\ref{eq:ScalarE2}).  We note that states of even (or odd) $J$ are
degenerate. This is unique among combinations of scalar and time-component
vector potential confinement \cite{ref:fifteen}.

It is important to note that the ``ultra-relativistic'' limit
where the quark mass vanishes is in fact not ultra-relativistic for scalar
confinement.  From the Hamiltonian (\ref{eq:Hscalar1}) with $p^2 = p_r^2 +
J^2/r^2$, the circular orbit condition is
\begin{equation}
{\partial H^2\over \partial r}\Bigg|_J = 0\ ,
\end{equation}
which implies a circular orbit radius of 
\begin{equation}
r_0^2 = {J\over a} \ .
\end{equation}
The circular velocity is then given by 
\begin{equation}
v_{\perp 0} = r_0 {\partial H \over \partial J}\Bigg|_{r=r_0} =
{1\over \sqrt2} \ .
\end{equation}
The massless quark moves at a velocity less than unity because the scalar
interaction contributes an effective mass of $ar_0$.

\subsection{Time-component vector confinement} 

The Hamiltonian (\ref{eq:Hvector}), with $m=0$ and the replacement $p^2 =
p_r^2 + {J^2\over r^2}$, becomes
\begin{eqnarray}
p_r^2 & = & (M-ar)^2 - {J^2\over r^2} \nonumber \\
& = & \left(M - ar - {J\over r}\right)\left(M - ar + {J\over r}\right)\ .
\end{eqnarray}

The first factor contains the classical turning points and the second has
only distant zeros.  To good approximation, we may use the zero condition
$M - ar = J/r$ from the first term in the second, and obtain
\begin{equation}
p_r^2 \simeq {2 J a\over r^2}\left(-r^2 + {M\over a} r - {J\over a}\right)\ .
\end{equation}
This is of the form of our general phase-space integrand in
Eq.~(\ref{eq:GenInt}) with $C=\sqrt{2Ja}$, $y=r$, where the turning points
satisfy 
\begin{eqnarray}
r_+ + r_- & = & {M\over a}\ , \nonumber \\
r_+ r_- & = & {J\over a} \ .
\end{eqnarray}

The quantization condition Eq.~(\ref{eq:GenSemiClass}) becomes
\begin{equation}
n + \frac12 = {\sqrt{2Ja}\over 2}\left[{M\over a} -2\sqrt{J\over
a}\,\,\right]\ .
\end{equation}
Solving for $M^2$, dropping the small squared radial excitation energy
and making Langer's replacement of $J$ by $J + \frac12$, we find
\begin{equation}\label{eq:vecsemiclass}
M^2 = 4 a\left(J + \sqrt2\, n + \frac12 + {1\over \sqrt2}\right)\ .
\end{equation}

\begin{figure}[htbp]
\epsfxsize = \linewidth
\hspace*{-5mm}
\epsfbox{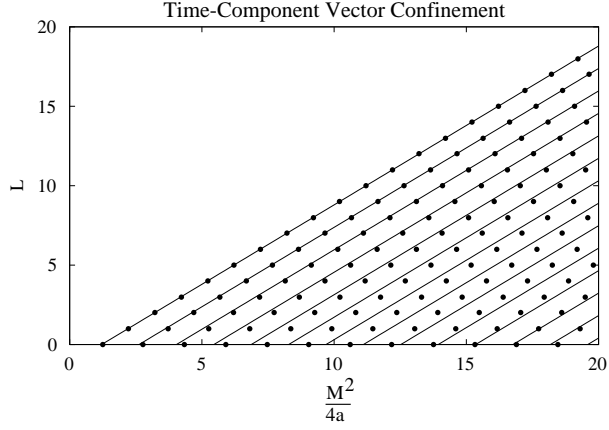}
\vskip 1 cm
\caption{Regge structure and states in pure linear time-component vector
confinement from numerical diagonalization of the Hamiltonian
in Eq.~(\protect\ref{eq:Hvector}) with $m=0$.  Solid lines are the approximate
semi-classical result of Eq.~(\protect\ref{eq:vecsemiclass}).}
\label{fig:two}
\end{figure}

Fig.~\ref{fig:two} shows the Regge spectrum of time component vector
confinement.  The semi-classical quantization method yields the correct
slope, radial excitation energy, and even nearly the correct $J=0$
intercept.

\subsection{Intermediate model}

\begin{figure}[htbp]
\epsfxsize = \linewidth
\hspace*{-5mm}
\epsfbox{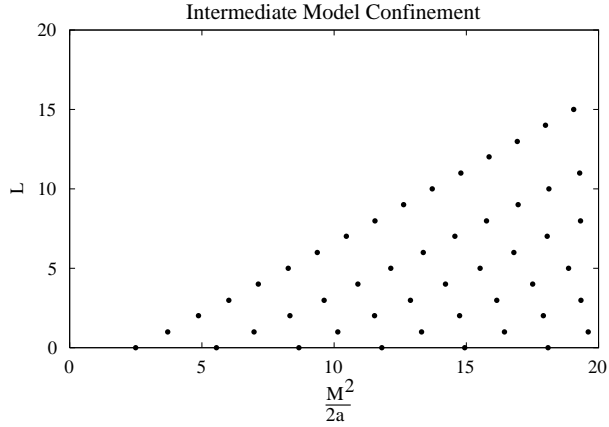}
\vskip 1 cm
\caption{States in pure intermediate model linear confinement from numerical
diagonalization of the Hamiltonian (\protect\ref{eq:Hint2}) with $m=0$.}
\label{fig:three}
\end{figure}

From the intermediate model Hamiltonian, Eq.~(\ref{eq:Hint2}), the Regge
spectrum can be exactly computed numerically, which we show in
Fig.~\ref{fig:three}.  The Regge trajectories are neither straight, nor
equally spaced.  The radial excitation energy is several times larger than
the scalar confinement potential.  A comparison of the intermediate and
scalar Hamiltonians reveals that they coincide in the classical circular
orbit limit.  It is in radial excitation that the two models differ
qualitatively.  Of course, even the quantized $n=0$ radial state has some
radial excitation.  A semi-classical quantization can also be done in this
case and yields a complicated transcendental relationship between $M^2$
and $J$.

\subsection{String confinement}

In this subsection we find that the Regge structure of the confining
string, with a massless quark at its end, resembles almost exactly scalar
confinement once the energy is rescaled.  This, despite the anomalous Regge
trajectories of the ``intermediate'' model which was supposed to mimic the
string.  We will later discuss the reason for the occurrence.  We begin
with the string Lagrangian (\ref{eq:fluxstring}).  The conserved quantities
$H$ and $J={\partial L\over \partial \omega}$ 
are \cite{ref:fourteen}
\begin{eqnarray}
{J\over r} & = & W_r\gamma_\perp v_\perp \nonumber \\
	& + & {ar\over 2v_\perp}\left({\arcsin
v_\perp\over v_\perp} - \sqrt{1-v_\perp^2}\right)\label{eq:stringJ} \  , \\
H & = & W_r \gamma_\perp  + ar {\arcsin v_\perp \over
v_\perp}\label{eq:stringH} \ , 
\end{eqnarray}
where the ``radial energy'' $W_r = \sqrt{ p_r^2 + m^2}$ was defined in
Eq.~(\ref{eq:Wr}) and $v_\perp = \omega r$. For circular orbits in the
massless quark limit the end of the string approaches the speed of light
($v_\perp\rightarrow 1$). Since this is the limit we are interested in for
the Regge structure we set $v_\perp = 1$ in the string quantities in
Eq.~(\ref{eq:stringJ}) and (\ref{eq:stringH}) to obtain
\begin{eqnarray}
{J\over r} & = & W_r\gamma_\perp v_\perp + {a\pi r\over
4}\label{eq:stringJ1} \ , \\ H & = & W_r \gamma_\perp +  {a\pi r\over
2}\label{eq:stringH1} \ ,
\end{eqnarray}

We do not set $v_\perp = 1$ in the quark terms since a delicate limiting
process occurs. In this limit, all of the angular momentum and energy
resides in the string and all of the radial momentum is carried by the
quark.

Next, we consider the difference of the squares of $H$ and ${J/r}$ 
\begin{equation}
H^2 - {J^2\over r^2} = W_r^2 + {a\pi r\over 2}W_r\gamma_\perp +
\frac34\left({a\pi r\over 4}\right)^2\ .
\end{equation}
Using Eq.~(\ref{eq:stringJ1}) to eliminate $W_r\gamma_\perp$, after a
little simplification we find
\begin{equation}
H^2 = p_r^2 + {J^2\over r^2} + {a\pi J\over 2} + \left({a\pi r\over
4}\right)^2\ , \label{eq:stringH2}
\end{equation}
where $W_r^2=p_r^2$ in the massless limit.  

If we define
\begin{eqnarray}
H_0^2 & = & H^2 - {a\pi J\over 2} \ , \\
a_0 & = & {\pi a\over 4}\ ,
\end{eqnarray}
the square of the string Hamiltonian appears to be a harmonic oscillator 
\begin{equation}\label{eq:stringHO}
H_0 ^2 = p_r^2 + {J^2\over r^2} + a_0^2 r^2\ , 
\end{equation}
which is very similar in form to the squared scalar confinement Hamiltonian
(\ref{eq:Hscalar1}).  

The squared string Hamiltonian in Eq.~(\ref{eq:stringH2}) has a critical 
difference from the harmonic
oscillator, as we now demonstrate. The circular orbit occurs where 
\begin{equation}
{\partial H^2\over\partial r}\Bigg|_J =  0 \ , 
\end{equation}
which implies that the circular orbit radius is 
\begin{equation}
r_0^2 = {4J\over a \pi}\ .
\end{equation}
The associated circular orbit velocity is 
\begin{equation}
v_{\perp 0} = r_0\,\omega = r_0 {\partial H\over\partial J}\Bigg|_{r=r_0} =
1 \ . 
\end{equation}

Thus, as we mentioned previously, the massless quark moves at the speed of
light in a circular orbit.  For radial excitation the quark moves in the
effective potential of Eq.~(\ref{eq:stringH2}).

\begin{figure}[htbp]
\epsfxsize = \linewidth
\hspace*{-5mm}
\epsfbox{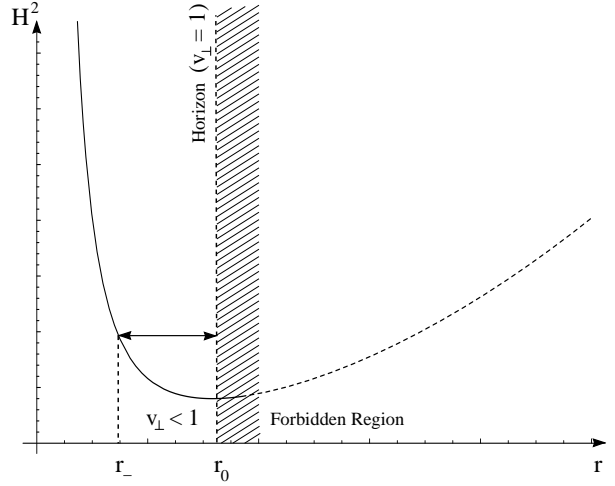}
\vskip 1 cm
\caption{Potential for the half-harmonic oscillator seen by a massless
quark on string, Eq.~(\protect\ref{eq:stringHO}).  The horizon is at the
minimum of the potential. The classical turning points are $r_-$ and
$r_0$.}
\label{fig:four}
\end{figure}

From the limiting form (\ref{eq:stringJ1}) of the angular momentum
(\ref{eq:stringJ}), we see that for radial motion the radius cannot exceed
$r_0$ because $W_r\gamma_\perp v_\perp$ cannot be negative. The $r=r_0$
coordinate represents a horizon or ``impenetrable barrier'' and the quark
moves in the ``half harmonic oscillator'' potential shown in
Fig.~\ref{fig:four}.

The semi-classical quantization of the string motion is equivalent to a
half harmonic oscillator shifted by an amount ${a\pi J\over 2} = 2 a_0 J$.
The half harmonic quantization condition is
\begin{equation}
\pi\left(n + \frac34\right) = {a_0\over 2}\int_{y_-}^{y_0} {dy\over
y}\,\sqrt{(y_+ - y)(y-y_-)} \ ,
\end{equation}
where $y = r^2$, $y_0 = r_0^2$, and $\Gamma = \frac34$, corresponding to
one smooth turning point.  The integral is not precisely one-half of the
full harmonic oscillator integral but the difference vanishes for large
$J$.  The result is
\begin{equation}
\pi\left(n+\frac34\right) = {\pi\over 8 a_0}\left[M_0^2 -
2a_0\left(J+\frac12\right)\right] \ ,
\end{equation}
or
\begin{equation}\label{eq:m02}
M_0^2 = 2 a_0 \left(J + 4n + \frac34 + \frac12 \right) \ .
\end{equation}
Finally, we rewrite Eq.~(\ref{eq:m02}) in terms of $M^2 = M_0^2 + {a\pi J
\over 2}$ and $a = {4 a_0 \over \pi}$ to obtain
\begin{equation}\label{eq:stringRegge}
M^2 = a \pi \left(J + 2n + \frac74\right) \ .
\end{equation}

We observe that the combination of the shift and the half oscillator
reproduces the $J + 2n$ pattern of excitation seen in the harmonic
oscillator, and hence in scalar confinement.

We can check the intercept ($J=0$) by directly quantizing the $s$-wave
states.  From Eq.~(\ref{eq:stringH}) with $\gamma_\perp = 1$, we have
\begin{equation}
H = p_r + ar \equiv M \ .
\end{equation}

\begin{figure}[htbp]
\epsfxsize = \linewidth
\hspace*{-5mm}
\epsfbox{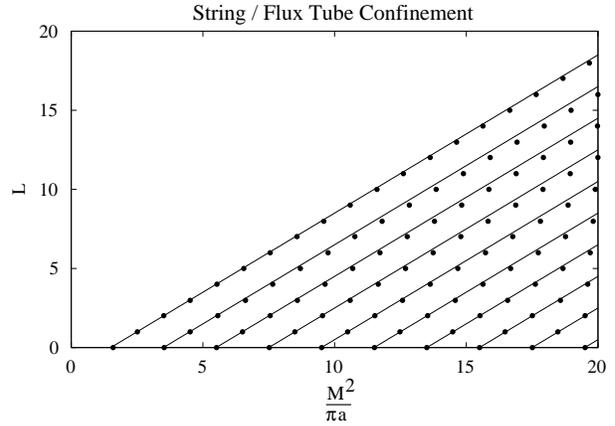}
\vskip 1 cm
\caption{Regge structure and states in string confinement from numerical
quantization of Eqs.~(\protect\ref{eq:stringJ}) and
(\protect\ref{eq:stringH}).  Solid lines are the approximate semi-classical
result of Eq.~(\protect\ref{eq:stringRegge}) with intercept $\frac32$ as
given in Eq.~(\protect\ref{eq:stringSwaveRegge}).}
\label{fig:five}
\end{figure}

The quantization integral,
\begin{equation}
\pi\left(n+\frac12\right) = \int_0^{M/a} dr\ (M - ar) = {M^2\over a} -
{M^2\over 2a}\ ,
\end{equation}
directly yields
\begin{equation}\label{eq:stringSwaveRegge}
M^2 = \pi a \left(2n + 1 + \frac12\right)\ ,
\end{equation}
where the $\frac12$ is the Langer correction for the radial equation.  The
result indicates the 3D harmonic oscillator.  We conclude that the true
intercept that should appear in Eq.~(\ref{eq:stringRegge}) ought to lie
between $\frac32$ and $\frac74$.  In Fig.~\ref{fig:five} we show the exact
numerical string Regge excitations with quark mass $m=0$.  The numerical
solutions of Eqs.~(\ref{eq:stringJ}) and (\ref{eq:stringH}) have been
discussed earlier \cite{ref:fourteen}.  The lines are the analytic solution
(\ref{eq:stringRegge}) but with intercept $\frac32$ from
Eq.~(\ref{eq:stringSwaveRegge}).  Similar solutions obtained from different
points of view have been obtained previously \cite{ref:sixteen}.

\section{Conclusions and Summary}\label{sec:six}

The concept of scalar confinement has been an important ingredient in
hadron model building for over two decades.  Its primary motivation was the
resulting pure Thomas type spin-orbit interaction which partially cancels
the vector type short range spin-orbit contributions.  Despite its
phenomenological success, scalar confinement has always had an uncertain
relationship with fundamental theory.  As pointed out by Buchm\"uller
\cite{ref:two}, the desired spin terms follow if the color magnetic field
vanishes in the quark rest frame.  This situation assumes no interaction
with the quark color magnetic moment and occurs naturally in the usual
color electric flux tube expected from QCD.  This observation originally
was proposed to justify the use of scalar confinement \cite{ref:two}.  We
emphasize here that this does not imply that the scalar potential follows
from QCD, only that they share a common spin-orbit interaction.

In this paper we have demonstrated that a four-vector confinement
interaction we found previously \cite{ref:eleven} is equivalent to 
scalar confinement.  This vector type interaction bears
a close resemblance to the QCD string, although there are significant
differences.  We have primarily considered here a class of confinement
models that share the same Thomas spin dependence.  Our comparison of
scalar and string/flux tube confinement has shown some interesting
differences and similarities even with spinless quarks.  We introduced an
intermediate model that has aspects of both scalar confinement and the QCD
string.  In this intermediate model the energy depends only on the
transverse quark velocity as expected in a straight string model.  The
interaction energy is effectively concentrated at the quark as in scalar
potential interaction.

The spin independent relativistic corrections of scalar and string
confinement differ, as has been known for some time
\cite{ref:five,ref:six}.  The relativistic corrections of the intermediate
model are as if an extra transverse mass $ar$ were concentrated at the
quarks position.  In the string case this same mass is distributed along
the string.

It is in the massless limit where interesting distinctions arise.  For pure
linear scalar confinement the meson mass is exactly given by $M^2 = 2a(J +
2n + 3/2)$, where $J$ and $n$ are the rotational and radial quantum
numbers.  The result, shown on the Regge plot in Fig.~\ref{fig:one}, is a
series of straight lines with an excitation pattern $J+2n$.  That is, there
are degenerate mass towers of states of even or odd parities.

The (laboratory frame) time-component vector confinement again produces
linear Regge trajectories, shown in Fig.~\ref{fig:two}, but with no tower
structure, owing to the excitation pattern $J+\sqrt2\, n$ with incommensurate
contributions from the rotational and radial quantum numbers.  Although one
might expect that QCD, being a vector interaction like QED, would have a
time-component interaction, it is evidently not time-component in the
laboratory frame.  This is precisely because the QCD field in which the
quark moves is not chromoelectrostatic (purely chromoelectric and
time-independent in the laboratory frame). Instead, the QCD field is
dynamical because the quark drags a chromoelectric flux tube along with it
as it moves. In this respect there are no ``test charges'' in QCD. The QCD
field is purely chromoelectric in its rest frame, leading to time-component
vector interaction in the quark's rest frame, which we have shown is
mathematically equivalent to a scalar interaction.  Neglect of the spatial
distribution of the QCD field energy thus leads directly to scalar
confinement.  The string/flux tube picture is the result of taking into
account the distribution of the field energy and momentum.

The intermediate model has a Regge structure very different from any of the
other models studied here, with somewhat curved trajectories and an uneven
pattern of radial excitation, as shown in Fig.~\ref{fig:three}.  Evidently,
the modification of the interaction that removes interaction contributions
to the radial momentum but leaves all the interaction energy and momentum
at the quark's position makes the intermediate model less, rather than
more, string-like in its consequences.

The string Regge spectroscopy, Fig.~\ref{fig:five}, again is similar to
that of scalar confinement, except with a different Regge slope.  Due to
the distribution of energy along the string, the quark now moves at the
speed of light in the massless limit.  This creates a horizon barrier so
the quark appears to move in a half oscillator.  The net effect is to give
an energy spectrum $M^2 = \pi a (J + 2n + 3/2)$ with the same tower of
states structure as in the scalar case.  Though the primary difference
between the two theories is the manner in which the energy and momentum of
the QCD field are distributed, the close relationship between their Regge
structures appears to be accidental.

We have pointed out a close, but not exact, relationship between scalar
confinement and the QCD string.  One might wonder whether one could change
the string tension and make the two even more similar.  The answer lies in
the static potential, which determines the low-lying states.  Presumably,
because they both arise from QCD, the string tension and the short-range
potential are correlated, as should be confirmed by experiment and lattice
simulation.  One cannot redefine the string tension arbitrarily.  Even were
scalar confinement and the string/flux tube to yield the same Regge slope,
their static potential and semi-relativistic reductions are different.

\section*{Acknowledgments}
This work was supported in part by the US Department of Energy under
Contract No.~DE-FG02-95ER40896.

\newpage
\vspace{2cm}

\vspace{2cm}

\vspace{2cm}

\vspace{2cm}

\end{document}